\documentclass[12pt,preprint]{aastex}

\newcommand{\chandra}{{\it Chandra}}
\newcommand{\asca}{{\it ASCA}}
\newcommand{\xmm}{{\it XMM-Newton}}

\newcommand{\kms}{km\,s$^{-1}$}

\newcommand{\nuth}{\nu_{\rm th}}
\newcommand{\stdplot}[1]{\includegraphics[angle=-90,width=3.5in]{#1}}
\newcommand{\tripspec}[1]{\includegraphics[angle=-90,width=3.0in]{#1}}
\newcommand{\nulll}[1]{}
\newcommand{\tnm}[1]{\tablenotemark{#1}}
\newcommand{\ud}[2]{$^{+#1}_{-#2}$}

\slugcomment{{\Large DRAFT: \today}}

\begin{document}

\title{Constraints on the Velocity and Spatial Distribution of Helium-like
Ions in the Wind of SMC X-1 from Observations with XMM-Newton/RGS}


\author{Patrick S. Wojdowski\altaffilmark{1}}

\affil{Kavli Institute for Astrophysics and Space Research,
  Massachusetts Institute of Technology} 

\email{pswoj@space.mit.edu}

\author{Duane A. Liedahl}

\affil{Department of Physics and Advanced Technologies, Lawrence
Livermore National Laboratory}

\author{Timothy R. Kallman}

\affil{NASA Goddard Space Flight Center}

\altaffiltext{1}{current address: Aret\'e Associates, P.O. Box 6024, Sherman Oaks, CA 91413} 

\begin{abstract}
We present here X-ray spectra of the HMXB SMC~X-1 obtained in an
observation with the \xmm{} observatory beginning before eclipse and
ending near the end of eclipse.  With the Reflection Grating
Spectrometers (RGS) on board \xmm{}, we observe emission lines from
hydrogen-like and helium-like ions of nitrogen, oxygen, neon,
magnesium, and silicon.  Though the resolution of the RGS is
sufficient to resolve the helium-like $n$=2$\to$1 emission into three
line components, only one of these components, the intercombination
line, is detected in our data.  The lack of flux in the forbidden
lines of the helium-like triplets is explained by pumping by
ultraviolet photons from the B0 star and, from this, we set an upper
limit on the distance of the emitting ions from the star.  The lack of
observable flux in the resonance lines of the helium-like triplets
indicate a lack of enhancement due to resonance line scattering and,
from this, we derive a new observational constraint on the
distribution of the wind in SMC~X-1 in velocity and coordinate space.
We find that the solid angle subtended by the volume containing the
helium-like ions at the neutron star multiplied by the velocity
dispersion of the helium-like ions must be less than
4$\pi$\,steradians\,\kms{}.  This constraint will be satisfied if the
helium-like ions are located primarily in clumps distributed
throughout the wind or in a thin layer along the surface of the B0
star.

\end{abstract}

\section{Introduction}

In isolated early-type stars winds are driven as ultraviolet photons
from the stellar surface impart their outward momentum to the wind in
resonance line transitions.  In a high-mass X-ray binary (HMXB), the
wind is ionized by X-rays from the compact object.  If the X-radiation
is intense enough, the resulting ions will not have transitions in the
ultraviolet and this greatly affects the dynamics of the wind
\citep[see, e.g., ][ and references therein]{blo94}.  The
SMC~X-1/Sk~160 system, which consists of a 0.71 second X-ray pulsar
and a B0I star together in a 3.9-day orbit, is the most X-ray luminous
known HMXB and, therefore, presumably, an extreme example of wind
disruption by X-ray ionization.

The behavior of the wind of an early-type star under the influence of
ionization from an X-ray emitting companion has been the subject of
many theoretical studies.  For SMC~X-1, the most relevant and detailed
such study is, arguably, the one by \citet{blo95} which included a
numerical hydrodynamic simulation.  The features that appeared in that
simulation included a regular, UV-driven wind on the X-ray shadowed
side of the star, a thermal wind on the X-ray illuminated side,
transition regions between the two types of winds and dense,
finger-like structures in the equatorial plane.  However, even this
simulation includes significant approximations: the gravity of the
compact object is not included and X-ray photoionization and its
dynamical effects are treated in a very approximate way.  Because of
the complexity of these systems and the difficulty of accounting for
all of the relevant physics, observations are critical to
characterizing the behavior of HMXB winds.

Though previous X-ray observations of SMC~X-1 have had spectral
resolving powers ($\equiv\lambda/\Delta\lambda=E/\Delta{}E$) of
$\sim$50 or less, X-ray spectroscopy has already revealed much about
the wind.  X-ray spectroscopic observations of this resolving power
can, through measurements of line fluxes, indicate the quantity and
ionization level of X-ray emitting material.  In observations of SMC
X-1 with \asca{}, emission lines were not detected and upper limits
were set on the quantity of material of intermediate ionization level
in the wind of SMC~X-1, excluding the presence of finger-like
structures as they appeared in the simulation of \citeauthor{blo95}
\citep{woj00}.  Further observations of SMC~X-1 with the Advanced
Camera for Imaging Spectroscopy (ACIS) on-board the {\it Chandra X-ray
Observatory}, however, have detected emission lines and, therefore,
the presence of material of intermediate ionization in the wind of
SMC~X-1 \citep{vrt01}.  These results indicate that overdense regions
do exist in the wind of SMC~X-1 and that while the simulations of
\citeauthor{blo95} may not be accurate in detail, the general form of
the structure predicted by \citeauthor{blo95} may, in fact, be
present.  To develop further constraints on models of the wind in
SMC~X-1, it is desirable to have constraints in addition to the
quantity and ionization level of the X-ray emitting plasma.
   
High-resolution spectroscopic observations with resolving powers
$\gtrsim$200 have the potential to provide useful constraints on the
kinematics of the high-ionization wind in HMXBs through direct
measurements of Doppler line shifts and broadening.  In addition, the
$n=2\to1$ triplets of He-like ions are resolved at high resolution and
measurements of the fluxes of the individual lines of these triplets
provide a constraint on the structure and kinematics of the X-ray
emitting material.  Of the three components of the helium-like
triplet, only one --- the resonance line --- has a large oscillator
strength and can be enhanced by resonant line scattering.  Because
resonant line scattering saturates and because this saturation depends
on the distribution of the scatterers in physical and velocity space,
constraints on the distribution of the wind in physical and velocity
space can be derived from the relative fluxes of these three lines
\citep{woj03}.  The flux ratios of the helium-like triplets are also
affected by ultraviolet radiation \citep*{blu72} and, therefore, the
flux ratios measured from an HMXB constrain the distance from the
photosphere of the high-mass star to the emission region.

Because of the location of SMC X-1 outside of the plane of the Galaxy,
the column density of interstellar material to it is small and, unlike
the HMXBs in the Galaxy, it can be observed in the wavelength range
15--35\,\AA.  The $n=2\to1$ transitions of the hydrogen-like and
helium-like ions of oxygen and nitrogen as well as hydrogen-like
carbon lie in this range.  Therefore, unlike with the Galactic HMXBs,
it is possible to derive observational constraints on the density and
velocity distributions of the regions of moderate ionization in
SMC~X-1 where these ions exist.  Because of the high spectral
resolution (FWHM of $\sim$65\,m\AA, corresponding to resolving powers
in the range 230--540) and large effective area (total of
$\sim$80\,cm$^2$) of the Reflection Grating Spectrometer on \xmm{} in
this band, the RGS on \xmm{} is particularly well-suited to
spectroscopic study there.  We present here observations of SMC~X-1
with \xmm{} beginning before an eclipse and ending near the end of
that eclipse.  We present an analysis in which we focus on the data
from the RGS and derive constraints on the space and velocity
distribution of material in the wind.  In \S\ref{sec:empir} we
describe our observations and an empirical analysis of the line
emission in which we measure line fluxes, shifts, and widths.  In
\S\ref{sec:phys} we derive constraints on the distribution of the wind
in physical and velocity space from the line fluxes we measure.  In
\S\ref{sec:discuss}, we discuss the implications of our results for
models of HMXB winds.

\section{Observations \& Empirical Analysis}
\label{sec:empir}

\begin{figure}
\includegraphics[angle=-90,width=5in]{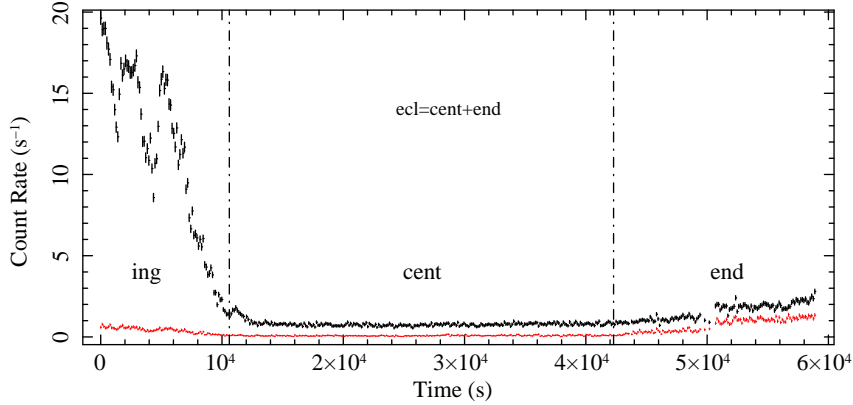}
\caption{Count rates from the MOS1 detector.  We show the count rate
  from a source region in black and from a background in red.  An
  increase at late time in both count rates is indicative of an
  increase in the non-X-ray background.  Each data point represents
  the average count rate over 128\,s.  Time zero on the abscissa is
  2001-05-31 02:22:16.9357 UT.}
\label{fig:m1_lc}
\end{figure}

SMC~X-1 was observed from on 2001-05-31 from approximately 02:14 to
18:50 UT with \xmm.  The actual times at which the first and last
observational data were received with the various instruments differ
from those times by as much as an hour.  However, data were collected
with each instrument 
during at least 90\% of the time between the nominal beginning and end
of the observations.  In Figure~\ref{fig:m1_lc} we show plots of count
rates from the source region and a background region on the MOS1
detector.
In the beginning, the count rate in the source region decreases
through X-ray eclipse ingress.  Toward the end of the observation,
the count rates for both the source and the background regions
increase while the difference stays approximately constant indicating
an increase in the non-X-ray background.  We divided the data into the
three time intervals noted on the figure.  On the time scale of the
figure, the times which divide the intervals are 10609\,s and
42289\,s.  The intervals, labeled ``ingress'', ``center'', and
``end'' were chosen to divide the eclipse ingress from the time of
complete X-ray eclipse and the time of high background from the rest
of the observation.  We then extracted spectra from each of the five
\xmm{} X-ray instruments for each of the three intervals and from the
interval consisting of the center and end intervals which we label
``eclipse''.

\begin{figure}
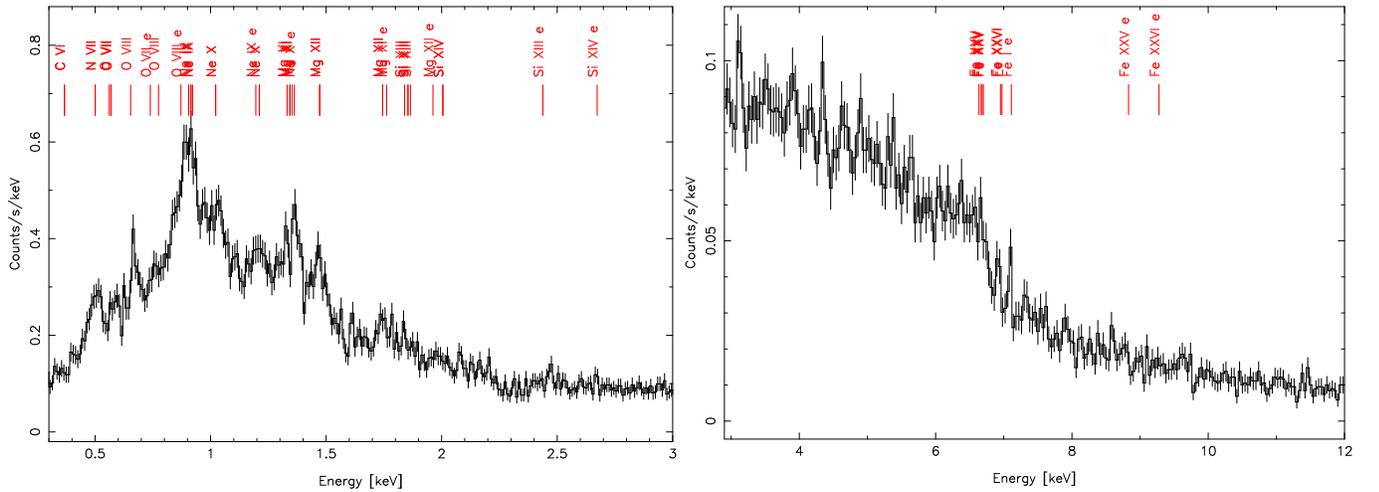

\stdplot{f2a.eps}
\stdplot{f2b.eps}
\caption{MOS 1 Eclipse spectrum of SMC X-1.  Labels indicate the
energies of spectral features which may or may not be detected in this
spectrum.  The labels denote the energies of lines except where the
ion is followed by ``e'' indicating an edge.  The line energies
indicated are $n=2\to1$ transitions except for \ion{0}{8} at
0.77\,keV, \ion{Ne}{10} at 1.21\,keV, and \ion{Mg}{12} at 1.74\,keV
which are $n=3\to1$ transitions.}
\label{fig:m1_ecl_spec}
\end{figure}

\begin{figure}
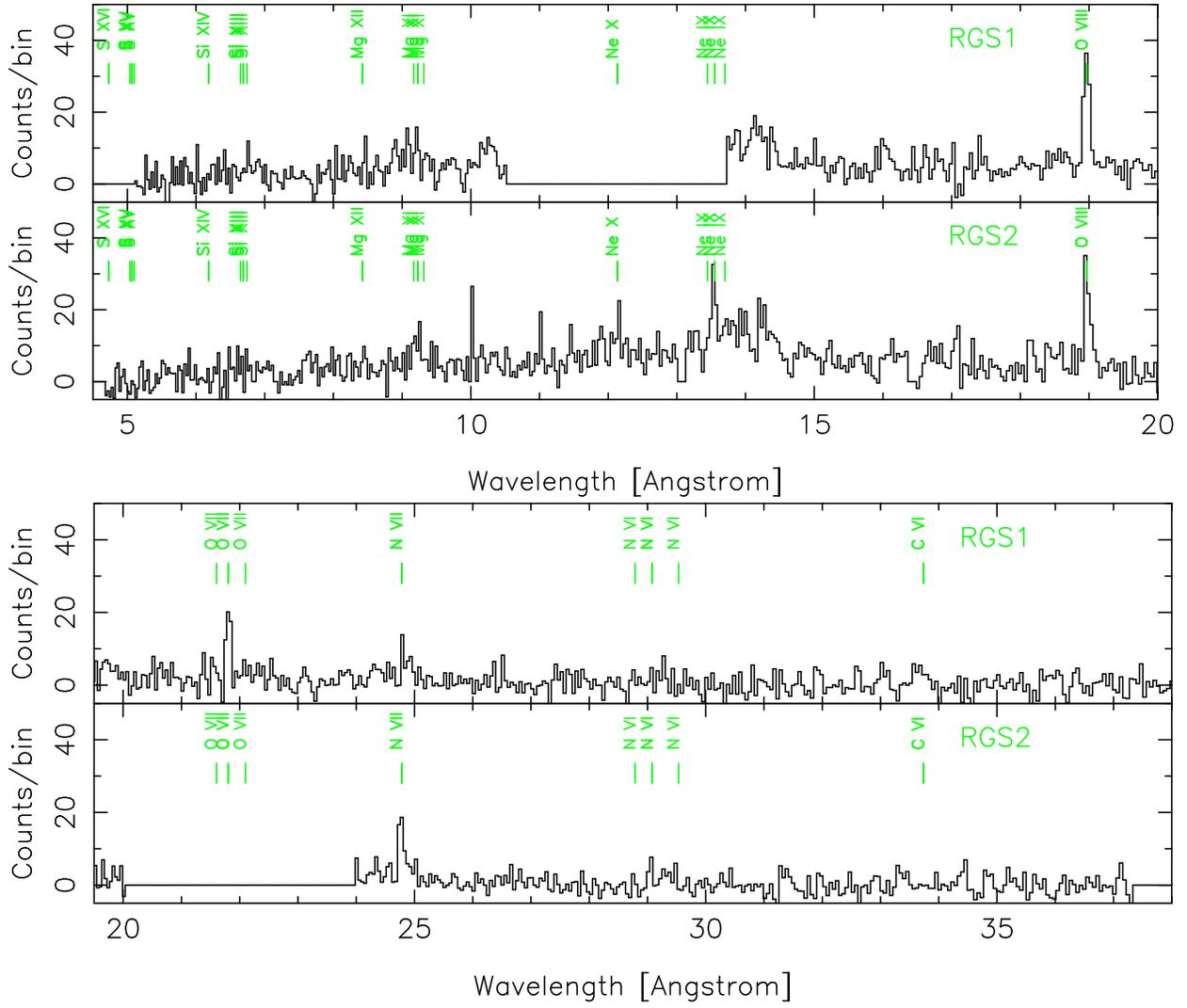

\noindent
\includegraphics[angle=-90,width=6.5in]{f3a.eps}
\includegraphics[angle=-90,width=6.5in]{f3b.eps}
\caption{The RGS eclipse spectrum of SMC~X-1.}
\label{fig:rgs_ecl}
\end{figure}

We focus here on the spectral data from the RGS.  However, in
Figure~\ref{fig:m1_ecl_spec} we plot the spectrum from MOS 1 for the eclipse
interval.
This spectrum is similar to those spectra of SMC~X-1 obtained by
\citet{vrt01} with \chandra/ACIS. In Figure~\ref{fig:rgs_ecl} we
show our RGS spectrum for the eclipse interval.
Line emission from hydrogen-like and helium-like ions of several
elements is apparent.  The spectral resolution of the \xmm{} RGS is
such that the $n=2\to1$ transitions of helium-like ions are resolved
into a triplets consisting of, in order of wavelength, the resonance
($1s2p\,^1\!P_1\to$ground), intercombination
($1s2p\,^3\!P_{2,1}\to$ground), and forbidden
($1s2s\,^3\!S_1\to$ground) lines.  However, in our RGS spectra, only
the intercombination lines are apparent.  Before attempting to explain
this fact or to use it to derive constraints on the properties of the
SMC~X-1 system, we first make a quantitative characterization of the
line emission.

In order to measure the fluxes, widths and shifts of these lines, we
fit the wavelength channels within 0.5\,\AA{} of the rest
wavelengths of each of the $n=2\to1$ emission line complexes.  We fit
the spectrum in the vicinity of each line complex using a power law
continuum and Gaussian lines.  We express the width (sigma) and
centroid of the Gaussians in velocity coordinates and fix the width
and shifts of the lines in a single complex to be the same, i.e.
\footnote{This awkward model for the continuum flux is due to the fact
that the power law spectral model is defined in energy space
($F_E\equiv{}K(E/{\rm keV})^\alpha$).  Our fits to the continuum
flux are not a focus of this work nor do we make any use of them
here so we proceed with this continuum model.}  :
\begin{equation}
F_\lambda=K(hc)^{\alpha+1}({\rm keV})^\alpha{}\lambda^{-(\alpha+2)}
+\sum_k\frac{1}{\lambda_k(v_r/c)\sqrt{\pi}} I_k 
\exp\left(\frac{(\lambda-\lambda_k(1+v_r/c))^2}{2\lambda_k^2(\sigma_{v,r}/c)^2}\right) 
\label{eqn:line_model}
\end{equation}
where $F_\lambda$ is the photon (not energy) flux, the $\lambda_k$s
are the rest wavelengths of the lines, and $K$, $\alpha$, $v_r$,
$\sigma_{v,r}$, and the $I_k$s are fit parameters.  We assume that
lines appear in emission and therefore constrain the fluxes of all
lines to be greater than or equal to zero.

\begin{figure}
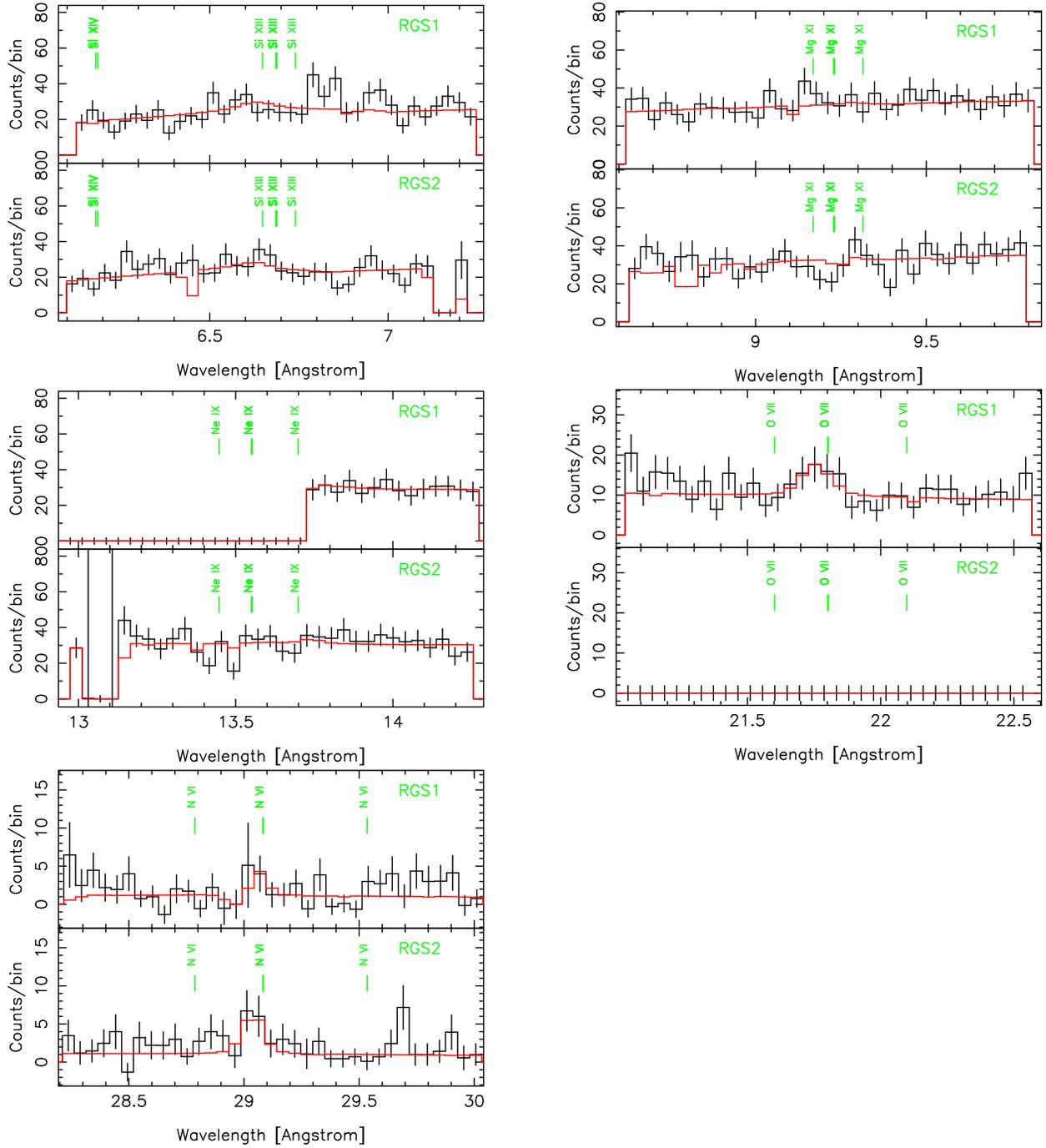

\tripspec{f4a.eps}
\tripspec{f4b.eps}
\tripspec{f4c.eps}
\tripspec{f4d.eps}
\tripspec{f4e.eps}
\caption{Helium-like triplets for the ``ingress'' interval.}
\label{fig:trips_ing}
\end{figure}
\begin{figure}
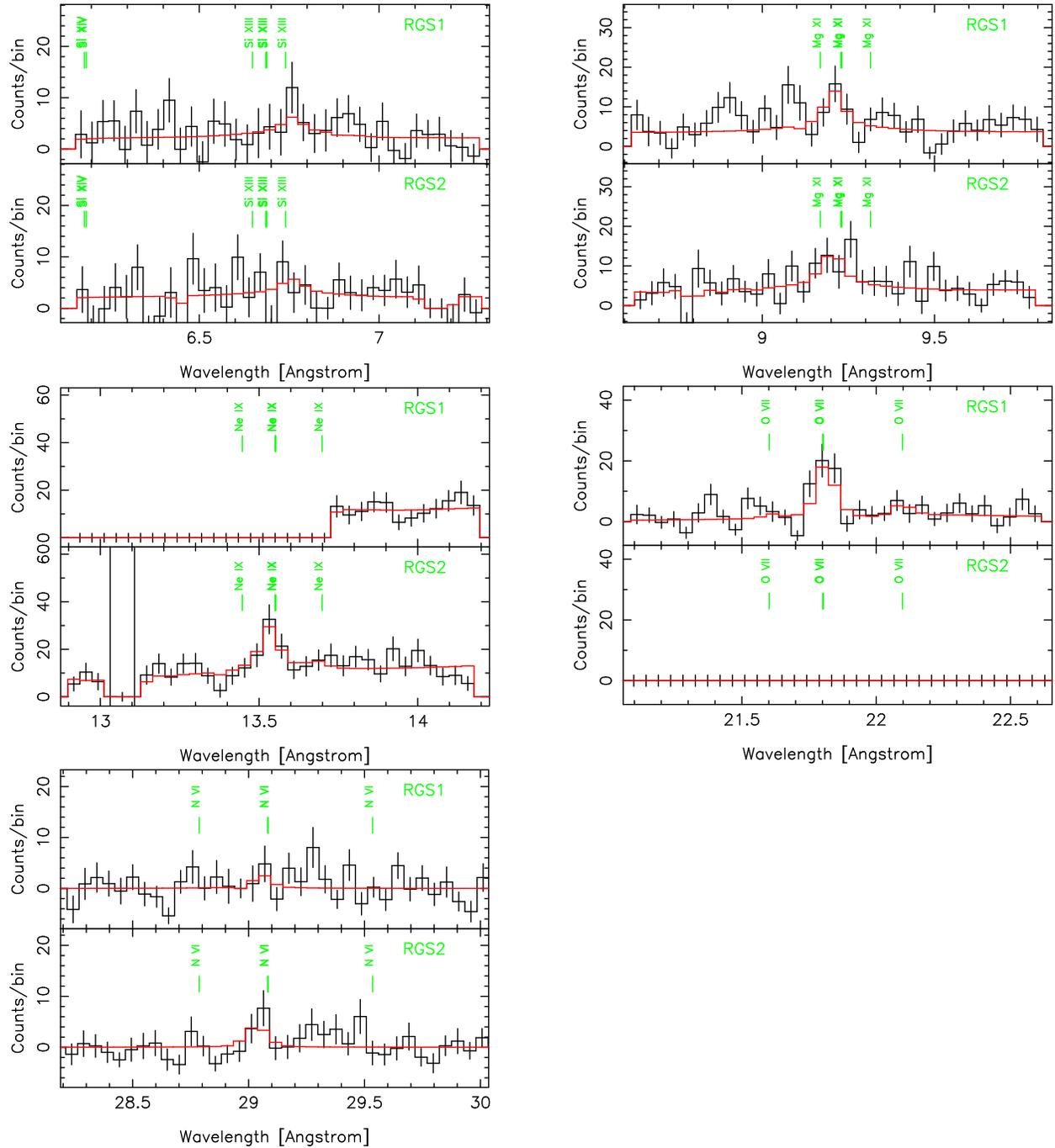

\tripspec{f5a.eps}
\tripspec{f5b.eps}
\tripspec{f5c.eps}
\tripspec{f5d.eps}
\tripspec{f5e.eps}
\caption{Helium-like triplets for the ``eclipse'' interval.}
\label{fig:trips_ecl}
\end{figure}

The intercombination line is actually an unresolved doublet.
Therefore, we fit the ingress and eclipse spectra in the vicinity of
the helium-like triplets using the model described by
equation~\ref{eqn:line_model} with four Gaussians: one for each of the
resonance and forbidden lines and one for each of the two components of
the intercombination line.  However, we fix the ratio of the fluxes of
the two components of the intercombination line to be equal to that
predicted by \citet{kin03} for recombination using temperatures given
in Table~\ref{tab:parconst}.  Furthermore, for reasons we describe
later, we formulate our model not in terms of the individual fluxes of
the three lines of the triplets but in terms of the total triplet line flux $I$ and
the ratios of the line fluxes $G$ and $R$ where
\begin{equation}
G\equiv\frac{I_{\rm i}+I_{\rm f}}{I_{\rm r}}
\end{equation}
and
\begin{equation}
R\equiv\frac{I_{\rm f}}{I_{\rm i}}
\end{equation}
where the subscripts ``r'', ``i'', and ``f'' indicate, respectively,
the resonance, intercombination, and forbidden lines.  In the fits, we
restrict the value of $\sigma_{v,r}$ to be less than 1500\,\kms{},
$v_r$ to be between $-$1000 and 1000\,\kms{} and we restrict $I$, $G$,
$R$, and $K$ to be greater
than or equal to zero.  In Figures~\ref{fig:trips_ing} and
\ref{fig:trips_ecl},
we show our best fits to these triplets and in
Table~\ref{tab:he_lines} we show the best-fit values and confidence
intervals for these line fluxes, shifts and widths.  The fact that we
detect only the intercombination lines of the helium-like triplets is
reflected in the fact that we obtain only lower limits on $G$ and only
upper limits on $R$.

\begin{deluxetable}{lccccccc}
\tablewidth{0pt}
\tablecaption{Measured Helium-like Triplet Parameters} 
\tablehead{ & & & & & & \multicolumn{2}{c}{power-law} \\
\cline{7-8}
Element & $I$\tnm{a} & $G$ & $R$ &
$\sigma_{v,r}$(\kms) & $v_{\rm r}$(\kms) & K & $\alpha$ } 
\startdata
\cutinhead{ingress}
N & 1.2\ud{0.7}{0.6} & $>$4 & \nodata & $<$800 & $-$300\ud{400}{300} & 
36\ud{10}{36} & 0\ud{\infty}{2} \\
O & 2.6\ud{1.4}{1.5} & $>$3& $<$0.3& 500\ud{600}{500} & $-$500\ud{600}{\infty}
& 135\ud{14}{111} & 0\ud{\infty}{3} \\
Ne & 1.4\ud{4.2}{1.4} & \nodata & \nodata & \nodata & \nodata & 170$\pm$30 &
$-$0.9\ud{\infty}{2.0} \\
Mg & 2\ud{6}{2}& \nodata& \nodata& \nodata& \nodata& 205\ud{16}{50} & $-$1.1\ud{1.1}{0.3} \\
Si & 22\ud{17}{16} & \nodata & \nodata & \nodata & \nodata & 280\ud{90}{70} & $-1.3\pm0.5$\\
\cutinhead{eclipse}
N & 0.22$\pm$0.18 & $>$1.6& $<$0.63& $<$500& $-$400\ud{500}{200} &
$<$1.3& \nodata \\
O & 1.1$\pm$0.4 & 18\ud{\infty}{14} & 0.26\ud{0.34}{0.26} & $<$300 &
90\ud{130}{270} & $<$0.0002 & 21.5\ud{0.6}{13.6} \\
Ne & 1.8\ud{0.7}{0.8} & $>$9& 0.2\ud{0.3}{0.2} & $<$600& $-$400\ud{500}{300}
& 6.5$\pm$0.7 & -6\ud{5}{\infty} \\
Mg & 2.2\ud{0.9}{0.7} & $>$2& $<$0.25& $<$1300& $-$600\ud{600}{\infty}
& 3.8\ud{1863.5}{0.6} & 0\ud{\infty}{3} \\
Si & 2.4\ud{2.4}{2.0} & \nodata & \nodata & \nodata & \nodata & 2.5\ud{1863.9}{0.8} 
& 0\ud{\infty}{9} \\
\enddata
\label{tab:he_lines}
\tablecomments{As a confidence limit on $\sigma_{v,r}$, $v_{\rm r}$, or
$\alpha$, $\infty$ indicates that parameter boundaries are within
the confidence limits.  The entry ``\nodata'' indicates that the data do
not constrain this quantity.}
\tablenotetext{a}{Total triplet line flux in units of $10^{-6}\,$photons$\,$cm$^{-2}$s$^{-1}$. Not corrected for interstellar photoelectric absorption.}
\end{deluxetable}

The hydrogen-like $n=2\to1$ transitions consist only of an unresolved
doublet and we fit these doublets using two Gaussians with the ratio of
the fluxes of those two lines fixed at 2:1, proportional to the
statistical weights of the upper levels and 
the ratio expected for recombination.
Furthermore, we restrict $\sigma_v$ to be less than 2500\,\kms{} and
$v_r$ to be between $-$2500 and 2500\,\kms.  In
Table~\ref{tab:h_lines}, we show the fit parameters for the
hydrogen-like lines.  Again, $I$ indicates the total photon flux for
the doublet.

\begin{deluxetable}{lccccccc}
\tablewidth{0pt}
\tablecaption{Measured Hydrogen-like Line parameters}
\tablehead{ & & & & \multicolumn{2}{c}{power-law} \\
\cline{5-6}
 $Z$ & $I$($10^{-6}$s$^{-1}$cm$^{-1}$) & $\sigma_{v,r}$(\kms) & 
$v_{\rm r}$(\kms) & norm & $\alpha$ }
\startdata
\cutinhead{ingress}
6 & 0.5\ud{0.6}{0.5} & \nodata & \nodata & $<$1900 &
\nodata \\ 
7 & 1.0\ud{0.7}{0.8} & $<$1200 & 600\ud{500}{1800} & 174\ud{17}{159} &
0\ud{\infty}{3} \\ 
8 & 4\ud{3}{2} & 900\ud{1100}{900} & $-$2300\ud{900}{\infty} &
16\ud{26}{13}& $-$6\ud{4}{\infty} \\ 
10 & 7$\pm$5 & 2500\ud{\infty}{2000} & \nodata& 171\ud{14}{13}
& 0.0\ud{\infty}{1.8} \\
12 & 1.4\ud{9.4}{1.4}& \nodata& \nodata& 170$\pm$30&
$-$0.4\ud{\infty}{0.3}  \\
14 & 44\ud{16}{22}& \nodata& 2000\ud{\infty}{4000}&
29000\ud{42000}{17000}& $-$8.6\ud{1.4}{1.2} \\ 
\cutinhead{eclipse}
6 & 0.3$\pm$0.3 & 1300\ud{\infty}{900} & $-$1500\ud{1900}{\infty}& $<$8
& \nodata \\  
7 & 0.7$\pm$0.2& $<$400& 0$\pm$200& 9$\pm$3& \nodata \\
8 & 1.7$\pm$0.3 & $<$400 & 160\ud{70}{140} & 8.4$\pm$1.9 &
0\ud{\infty}{3} \\ 
10 & 1.0$\pm$0.6& \nodata& 200\ud{900}{1100}& 7.5$\pm$1.4 & 0\ud{\infty}{5}
\\
12 & 0.9\ud{1.0}{0.9}& \nodata & \nodata & 17\ud{1892}{4}& \nodata \\ 
14 & $<$2.0 & \nodata & \nodata & 3.3\ud{2632.6}{1.6} & \nodata \\ 
\enddata
\tablecomments{The entry ``\nodata'' indicates that the data do not
constrain this quantity.}
\label{tab:h_lines}
\end{deluxetable}

\section{Line Fluxes as Measures of the Distribution of Scatters in
  Physical and Velocity Space}
\label{sec:phys}

The emission lines from high-mass X-ray binaries result from
recombination and resonant line scattering in wind of the high-mass
star.  The line emission from these processes depends on the
distribution of the wind material in coordinate space and, in the case
of resonant line scattering, in velocity space.  Therefore, the line
fluxes we measure provide a constraint on these distributions of the
wind material.  However, the equations that describe these
distributions are complex and rather than attempt to derive
constraints directly from the line fluxes, we believe it will be more
illuminating to create a simple model described by a few parameters
and then derive constraints on these parameters.  In \S\ref{sec:rnr},
we derive equations for line luminosities due to recombination and
resonant line scattering.  In \S\ref{sec:helike}, we describe a
modification to those luminosities for the case of the helium-like
triplets due to ultraviolet radiation of the high-mass star.  In
\S\ref{sec:models} we describe the simple models and our constraints
on their parameters.

\subsection{Recombination and Resonant Line Scattering}
\label{sec:rnr}

In a plasma in photoionization equilibrium, photons are absorbed from
the continuum in photoionizing transitions and line photons are emitted
as electrons and ions recombine.  If the plasma is optically thin in
the photoionizing continuum, the recombination line luminosity may be
expressed as
\begin{equation}
L_{{\rm rec}} =  h\nu_0\frac{\pi{}e^2}{m_ec}
(4\pi)^{-1}\int\!\!\!\int \eta n_{{\rm Ion}}ds\,d\Omega
\int_{\nuth}^\infty\frac{L_\nu}{4\pi{}h\nu}\frac{df}{d\nu}d\nu 
\label{eqn:lrec}
\end{equation}
where $h$ is Planck's constant, $\nu_0$ is the frequency of the line,
$e$ and $m_e$ are the electronic charge and mass, $n_{{\rm Ion}}$ is
the density of the ion that emits the line, $\eta$ is fraction of
recombinations to the ion which result in emission of the line,
$L_\nu$ is the specific luminosity of the compact X-ray source,
$\nuth$ is the ionization threshold frequency of the emitting ion in
its ground state, and $df/d\nu$ is its continuum oscillator strength,
and $s$ and $\Omega$ are the distance from and solid angle subtended
at the compact X-ray source.

In any plasma, line photons are absorbed and reemitted in resonance
line transitions.  The net effect of this is only to redirect line
photons.  However, depending the geometry of the plasma and radiation
source, resonant line scattering may result in absorption or emission
lines being observed.  For the geometry of interest in this work --- an
occulted compact radiation source with radiation scattered into the
observer's line of sight by plasma --- resonant scattering results in
emission lines and the line luminosity due to scattering may be
expressed as
\begin{equation}
L_{{\rm scat}} = L_{\nu_0}(4\pi)^{-1}\int\!\!\!\int
(1-e^{-\tau(\nu,{\bf \Omega})})d\nu{}d\Omega 
\label{eqn:lscat}
\end{equation}
where $\tau$ is the line optical depth through the plasma along a line
from the compact radiation source.  The line optical depth is given by
\begin{equation}
\tau=f\frac{\pi{}e^2}{m_ec}\int{}n_{\rm Ion}\phi(\nu)ds
\end{equation}
where $f$ is the line oscillator strength and the function
$\phi$ describes the line profile.  Neglecting natural line
broadening and considering only Doppler line broadening, we have
\begin{equation}
\phi(\nu)\equiv\frac{c}{\nu_0}g(c(\nu/\nu_0-1))
\end{equation}
where $g$ is the local distribution of the ions' velocity component in
the direction along the line of sight from the compact X-ray source
and is normalized such that
\begin{equation}
\int_{-\infty}^{\infty}g(v)dv\equiv1. 
\end{equation}
However, we assume that ion velocities are very small compared to $c$
so that $g(v)$ is non-zero only for $|v|\ll{}c$.  With this definition,
\begin{equation}
\int_0^{\infty}\phi(\nu)d\nu=1.
\end{equation}

As we will ultimately be interested in calculating observed line
fluxes the spatial integrals in equations~\ref{eqn:lrec} and
\ref{eqn:lscat} should be restricted to the volume that is both
illuminated by the compact radiation source and observable (i.e., not
occulted by the star).  In principle, the radiation from the compact
source that reaches the observable plasma may be depleted at the line
frequency by line scattering in the occulted region, reducing the
observable line luminosity relative to that described by
equation~\ref{eqn:lscat}.  However, as scattering does not destroy
photons but merely redirects them, for most plasma geometries, the
total number of line photons reaching the observable region will not
be significantly affected by scattering in the occulted region.  Only
for a specialized plasma geometry: a long, narrow distribution of
plasma, oriented along a line of sight from the compact radiation
source will equation~\ref{eqn:lscat} be significantly inaccurate.
Furthermore, if the part of the wind where the line emission occurs is
moving supersonically (as HMXB winds are known to do) the line
frequencies in different parts of the wind will be shifted apart so
that resonant line scattering in one part of the wind will not be
affected by resonant line scattering in another part.

In equation~\ref{eqn:lrec}, we have $\eta$ inside
the spatial integral, allowing for the possibility that the
recombination efficiencies may depend on the local conditions.  In
fact, for helium-like lines, the forbidden line
may be ``pumped'' into the intercombination line at high densities or
high ultraviolet radiation intensities and we define $\eta$ so that
the effects of this pumping are absorbed into it.  Near a hot star such as
Sk~160, the B0 companion of SMC~X-1, the ultraviolet radiation
intensity is high.  Therefore, the line luminosities due to resonant
scattering depend on the distance from the star.  In the next section,
we describe this in more detail.

\subsection{Helium-like Triplets}
\label{sec:helike}
As can be seen from the equations above, the observed flux of any one
line depends in a complicated way on the distribution of the emitting
ions in both coordinate and velocity space.  Therefore, the constraint
that can be derived from a single line is quite complex.  However,
because the resonance line of each helium-like triplet has a
significant oscillator strength while the other two lines do not,
only the resonance line is affected by resonant scattering.
Therefore, only the flux of the resonance line depends on the velocity
distribution of the emitting ions and by measuring the flux of all
three lines of the triplet, it is possible to derive more specific
constraints on the distribution of the emitting ions in space and
velocity than can be obtained from the flux of any one line.



As previously mentioned, the recombination efficiencies of the
intercombination and forbidden lines are not constant but depend on
the local conditions.  This is due to the fact that the upper level of
the forbidden line, the $^3S_1$ level, is metastable and an ion in
that state may be excited by absorption of an ultraviolet photon or
collision with an electron to any of the $^3P$ levels.  Two of these
levels ($^3P_2$ and $^3P_1$) make up the upper levels of
the intercombination line.  Therefore, at high densities or high
ultraviolet radiation intensities, the effective recombination
efficiency of the intercombination line is increased and the effective
recombination efficiency of the forbidden line is decreased.  This
effect has been used by several authors to estimate the distance of
emitting ions from isolated hot stars.  However, several of the
calculations have neglected the fact that once an ion is excited from the
$^3S_1$ state to a $^3P$ state, it may decay back to the $^3S_1$ state by
emission of an ultraviolet photon rather than decay to ground by
emission of an intercombination line photon.  We write the
recombination line efficiencies including this effect as 
\begin{eqnarray}
\eta_{\rm i}&=&\eta_{\rm i}^\prime+\eta_{\rm f}^\prime(1-b_{\rm f}) \\
\eta_{\rm f}&=&\eta_{\rm f}^\prime{}b_{\rm f}
\end{eqnarray}
where $\eta_{\rm i}^\prime$ and $\eta_{\rm f}^\prime$ are the
recombination efficiencies without $^3S_1\to{}^3P$ excitation and
$b_{\rm f}$ is the branching ratio for radiative decay of the $^3S_1$
state to ground.  In order to derive an expression for this branching
ratio, we solve the rate equations of \citet[][ equations 18 and
19]{mew78} and neglect inner-shell ionization, all collisions, and
induced radiative decay.  To write our expression, we use the notation
of \citet{por01} and denote the $^3S_1$ level by $m$ (for
``metastable''), the ground ($1s^2\,^1S_0$) by $g$ and denote the
level $^3P_k$ by $p_k$ where $k$ may be 0, 1, or 2 and get
\begin{equation}
b_{\rm f}\equiv\frac{A_{mg}}{A_{mg}+T}
\end{equation}
where 
\begin{equation}
T\equiv\sum_{k=0}^2w_{mp_k}\frac{A_{p_kg}}{A_{p_kg}+A_{p_km}}
\end{equation}
where $w_{mp_k}$ is the excitation rate from the $^3S_1$ to
$^3P_k$\footnote{Our expression for the forbidden line efficiency is
consistent with expressions for forbidden line intensities given by
\citet{mew78} and by \citet{por01}.  However, our expression for the
intercombination line is not consistent with those works.  As they are
written, it is difficult to compare our line efficiencies with those
line intensities.  Therefore, we note that our solution of the rate
equations implies
$I_i=\sum_{k=1}^2C_{gp_k}(1-BR_{p_km}BR_{mg})+C_{gm}(1-BR_{mg})$ for
equation 5 of \citet{por01}.  We are encouraged in using our
expression for the intercombination line because, with our expression,
the sum of the intercombination and forbidden lines, as expected, does
not depend on the magnitude of the $^3S_1\to^3P$ excitation.  In any
case, the final results for both of those works are based not on those
expressions but on solutions of the full rate matrix and would not be
affected by any error in the expressions.}.

For photoexcitation, the $w$ values are given by
\begin{equation}
w_{mp_k}=\frac{4\pi{}^2e^2}{m_ec}f_{mp_k}\frac{J_{\nu_{mp_k}}}{h\nu_{mp_k}}
\end{equation}
where $J_{\nu_{mp_k}}$ is the mean intensity of radiation (i.e., the
intensity averaged over solid angle) at the frequency $\nu_{mp_k}$ and
$\nu_{mp_k}$ and $f_{mp_k}$ are, respectively, the frequency and
oscillator strength of the transition $m\to{}p_k$.  The frequencies
$\nu_{mp_k}$ for elements from carbon to iron range from the near to
the extreme ultraviolet.  Because the quantity $\eta_{\rm i}+\eta_{\rm
  f}$ and the value of $\eta_{\rm r}$ do not depend on the ultraviolet
pumping, neither does the value of the $G$ ratio but the value of the
$R$ ratio does.

\subsection{Simple Model}
\label{sec:models}
From the observed fluxes of the three lines of a given helium-like
triplet, it is possible to derive constraints on the distribution of
the helium-like ions in coordinate and velocity space.  However,
because of the complexity of the equations describing the line
luminosities, the constraints that can be derived are also complex.
Therefore, we compute line fluxes for two simple parameterized models
for the distribution of material.  Then we derive constraints on those
parameters from the observed line fluxes.  We do not actually expect
the distribution of material in the wind of SMC~X-1 to be described by
these simple models in detail.  However, from these model parameter
constraints it is possible to make approximate inferences about the
distribution of the wind of SMC~X-1 in coordinate and velocity space.

In our models, each ion exists exclusively within a solid angle
$\Omega$ subtended at the neutron star, and is distributed such that
the column density along lines of sight from the neutron star has the
single value $N$ within the solid angle $\Omega$.  The distribution of
ion velocities along lines of sight from the neutron star in our model
is uniform over some range with width $\Delta{}v$ (which need not be
centered on zero).  Also, to simplify the calculation of ultraviolet
pumping, we take the entire plasma to be at a single distance $r$ from
the surface of the companion star which we take to be a sphere
emitting as a blackbody with a temperature of $T_\star=$30,000\,K, the
approximate effective temperature for a star of type B0 such as
Sk~160.

For this model, the UV radiation density is
\begin{equation}
J_{\nu}=W(r)B_\nu(T_\star)
\end{equation}
where $B_\nu$ is the Planck function,
\begin{equation}
W(r)\equiv\frac{1}{2}
\left[1-\left(1-\left(\frac{r_\star}{r}\right)^2\right)^{1/2}\right],
\end{equation}
and $r_\star$ is the stellar radius.  The luminosity of a line due to
recombination is then
\begin{equation}
L_{{\rm rec}}=h\nu_0\frac{\pi{}e^2}{mc}
(4\pi)^{-1}\eta\Omega{}N 
\int_{\nuth}^\infty\frac{L_\nu}{4\pi{}h\nu}\frac{df}{d\nu}d\nu 
\end{equation}
and the line luminosity due to scattering is
\begin{equation}
L_{{\rm scat}}=L_{\nu_0}(4\pi)^{-1}\frac{\nu_0}{c}\Omega\Delta{}v
(1-e^{-f\frac{\pi{}e^2}{m\nu_0}\frac{N}{\Delta{}v}}).
\label{eqn:lscat_model}
\end{equation}
or, equivalently,
\begin{equation}
L_{{\rm scat}}=L_{\nu_0}(4\pi)^{-1}\frac{\nu_0}{c}\Omega{}N
\frac{1-e^{-f\frac{\pi{}e^2}{m\nu_0}\frac{N}{\Delta{}v}}}{N/\Delta{}v}.
\label{eqn:lscat_model2}
\end{equation}

It can be seen from the above expressions that it is not necessary to
specify all of the model parameters in order to determine the line
luminosities.  It is sufficient to specify only the parameter
combinations $\Omega{}N$ and $N/\Delta{}v$ rather than the values of
the three parameters contained in those two combinations.  If we replace the
three parameters with the two parameter combinations then the fluxes
of the lines depend linearly on $\Omega{}N$, the sum of the
luminosities of the intercombination and forbidden lines depends only
on $\Omega{}N$, the $G$ ratio depends only on $N/\Delta{}v$, and the
$R$ ratio depends only on $r$.

In principle, it should be possible to derive constraints on the model
parameters from the constraints on the line fluxes in
Table~\ref{tab:he_lines}.  However, for the line emission mechanisms
we consider here, the $G$ and $R$ ratios can only take on values
within a finite range (e.g., see \citealt{woj03} Figure~8 for the
range of the $G$ ratio for \ion{Si}{13}).  In at least one case, our
best fit value of $G$ falls outside of this range.  Therefore, in
order to derive meaningful constraints on the physical model
parameters, we redo the fits with the $G$ and $R$ values constrained
to be within the range allowed by scattering and recombination as
described above.  In all cases, we are able to obtain good fits with
these constraints imposed.

In order to relate our model parameters to line fluxes and derive
constraints on our model parameters, we use the following data.  We
get recombination efficiencies (without $^3S_1\to^3P$ pumping) by dividing
effective line recombination rates from \citet{kin03} by total
recombination rates from \citet{ver96c}.  We have written equations
for line luminosities in terms of the specific luminosity of the X-ray
source.  Of course, we measure fluxes and any luminosities we derive
are subject to uncertainty in the distance to the object.  However,
if we write equations~\ref{eqn:lrec} and \ref{eqn:lscat} in terms of
fluxes, factors of distance cancel.  Therefore, in our calculations we
use fluxes and our results do not depend on the distance to SMC~X-1.
For the specific luminosity of the SMC~X-1 neutron star $L_\nu$ we use
\begin{equation}
{\cal F}=K\left(\frac{\epsilon}{1\,{\rm keV}}\right)^{-\alpha}
\end{equation}
where ${\cal F}$ is the specific photon flux per unit photon energy
interval which is related to $L_\nu$ by
$L_\nu=(4\pi{}d^2)^{-1}h^2\nu{\cal F}$ and $\epsilon$ is photon energy with
$K=0.0267$\,photons\,cm$^{-2}$\,keV$^{-1}$ and $\alpha=1$ from a fit
to the uneclipsed {\it ASCA} spectrum of SMC~X-1 \citep{woj00}.  The
value of 1 we use for $\alpha$ is an approximation to the best-fit
value of 0.94$\pm$0.02.  
For the continuum oscillator
strength of the helium-like ions above the ionization threshold we use
\begin{equation}
\frac{df}{d\nu}=2\nuth^2\nu^{-3}
\end{equation}
\citep[c.f.,][]{woj03}.  For all elements, we take the oscillator
strength of the resonance line to be 0.7, which is accurate to 5\%.
For line frequencies (wavelengths) and spontaneous decay rates we use
values from the Astrophysical Plasma Emission Database
\citep[APED,][]{smi01}.  We take the oscillator strengths $f_{mp_0}$ to
be zero and derive the oscillator strengths $f_{mp_1}$ and $f_{mp_2}$
from the spontaneous deexcitation rates of APED.  We give the results
of our model parameter constraints in Table~\ref{tab:parconst}.

\begin{deluxetable}{lccccccccccc}
\tablewidth{0pt}
\tablecaption{Simple Model Parameters Derived from Helium-like Line Emission}
\tabletypesize{\scriptsize}
\tablehead{element & $T_{\rm rec}$(eV) & $\Delta\tau$ &
$r/r_\star$\tnm{a} &
$N/\Delta{}v$ & $\Omega(4\pi)^{-1}N$ & $\Omega(4\pi)^{-1}\Delta{}v$ &
$T_{\rm rec}$ \\
 & & &  &
($10^{15}$\,cm$^{-2}$(\kms)$^{-1}$) &
($10^{15}$\,cm$^{-2}$) & (\kms) & reference}
\startdata
N  & 3.0 & $>$4.2  & $<$57         & $>1.0$  &
0.22$\pm$0.18    & $<$0.39 & 1 \\
O  & 4.0 & $>$20   & 18\ud{11}{17} & $>6.4$  & 1.3$\pm$0.4      & $<$0.27 & 1 \\
Ne & 12  & $>$22   & $<$6.6        & $>11$   &
2.2\ud{0.9}{1.0} & $<$0.27 & 2 \\ 
Mg & 21  & $>$4.2  & $<$1.4        & $>3.2$  &
3.4\ud{1.4}{1.1} & $<$1.5  & 2 \\
Si & 75  & \nodata    & \nodata    & 5.1\ud{5.1}{4.2} & 
\nodata{} & \nodata & 2 \\
\enddata
\tablenotetext{a}{$r$ is the distance from center of the star.}
\label{tab:parconst}
\tablerefs{(1)~\citealt{kin03}; (2)~\citealt{sak99}}
\end{deluxetable}

\section{Summary/Discussion}
\label{sec:discuss}

We have observed SMC~X-1 in eclipse with \xmm{} and, with the RGS,
resolved the helium-like triplets of nitrogen, oxygen, neon, and
magnesium.  To our knowledge, this is the first time helium-like
triplets from SMC~X-1 have been resolved and the first time the
helium-like triplets of nitrogen, oxygen, or neon have been resolved
for any HMXB.  In all cases, only one component of the triplet, the
intercombination line, is observably present.  The lack of observable
forbidden line fluxes in the triplet is easily explained by
photoexcitation pumping of the forbidden line into the
intercombination line by the ultraviolet radiation of the B0 star
Sk~160 and allows upper limits to be set on distance of the emitting
helium-like ions from Sk~160.  The absence of observable fluxes in the
resonance lines is consistent with what we expect from recombination
in the photoionized wind.  However, the fact that the resonance lines
are not enhanced by resonant scattering implies a lower limit on the
optical depth of the wind in the resonance line and, therefore,
constraints on the structure and kinematics of the wind are also
implied.

We have modeled the helium-like line emission as recombination and
scattering emission from a region that is partly obscured by the
companion star and is described by a single solid angle ($\Omega$) and
column density ($N$) and the velocities along lines of sight from
the neutron star have a boxcar distribution with width $\Delta{}v$ for
each of the ions.  We have inferred
infer the quantity $\Omega\Delta{}N$ from the sum of the flux of the
intercombination and forbidden lines which are not affected by
resonant line scattering.  We infer the quantity
$N/\Delta{}v$ from flux of the resonance line relative to
the sum of the other two (the inverse of the $G$ ratio).  From the
constraints on these two quantities, we also obtain a constraint on
the quantity $\Omega\Delta{}v$.  Because we do not detect the
resonance line, we obtain only lower limits on $N/\Delta{}v$
and only upper limits on $\Omega{}\Delta{}v$.  

As previously mentioned, in our analysis we have assumed that photons
escape isotropically along lines from the location of their production
or first scattering.  While we do expect recombination emission to be
isotropic, resonant line scattering is not isotropic.  The angular
distribution for resonant line scattering is a linear combination of
an isotropic distribution and an dipole distribution \citep[see,
e.g.,][Ch.\ 19]{cha60}.  Furthermore, for large optical depths that
are comparable to or greater than unity such as we infer, photons may
undergo several scatterings before escaping.  In this case, photons
will escape preferentially along directions where the optical depth is
least.  In general, the calculation of photon escape direction is also
complicated by the fact that photons may travel large distances across
the plasma before escaping.  While it is difficult to include this
possibility in analytic calculations, it is straightforward using
Monte Carlo techniques.  However, if the region that emits the
helium-like line emission has supersonic velocity differentials, as we
know that the bulk of the wind does, the transfer of resonant line
photons is essentially local (the Sobolev approximation,
\citealt{cas70}) and can be approached analytically.  For a
supersonically expanding wind, the optical depth in a direction given
by the coordinate $x$ is inversely proportional to $dv_x/dx$.  If the
flow moves along lines from a single point and $y$ is the distance
from that point, then these derivatives, in directions parallel and
perpendicular to the flow, respectively, are $dv/dy$ and $v/y$.  While
these quantities will not generally be equal, they will generally be
of the same order over the bulk of the wind and, therefore, if we
assume complete redistribution in frequency and direction for
individual scatterings, then the radiation will escape isotropically
\citep[see][]{cas70}.  In fact, individual scattering events do not
redistribute frequency and direction completely and so even with
$v/s=dv/ds$, radiation does not escape isotropically.  However the
difference from the case of complete redistribution is no greater than
10\%--20\% (\citealt{car72}, see also \citealt{mih80}).  In light of
these facts, we proceed to explore the implications of our results.
However, it must be kept in mind that the approximations we have made
for the radiative line transfer may quantitatively affect our
conclusions.

The implication of our results on fundamental wind parameters, such as
the mass-loss rate and the terminal wind velocity, is complex owing to
factors such as the complex dependence of the ion fractions on density
in photoionization balance.  Therefore, interpreting these results in
terms of fundamental wind parameters is difficult to do without
computing line fluxes for complete wind models which is beyond the
scope of this work.  However, the upper limits on
$(\Omega/4\pi)\Delta{}v$ that we obtain, approximately 1\,\kms, are
quite small compared to 1000\,\kms{}, the order of the terminal wind
velocities of massive stars and the wind velocity in a simulation of
the wind of SMC~X-1 by \citet{blo95}.  Therefore, the part of the wind
that emits the helium-like lines cannot be homogeneously distributed
around the entire wind.  Instead, we believe that our results indicate
either that the volume (or volumes) containing the helium-like ions of
the various elements subtend a small solid angle at the neutron star
or that the helium-like ions exist primarily in a region where the
wind is at a small fraction of its terminal velocity or, perhaps, some
combination of both.

\begin{figure}
\begin{center}
\includegraphics[angle=-90,width=2.4in]{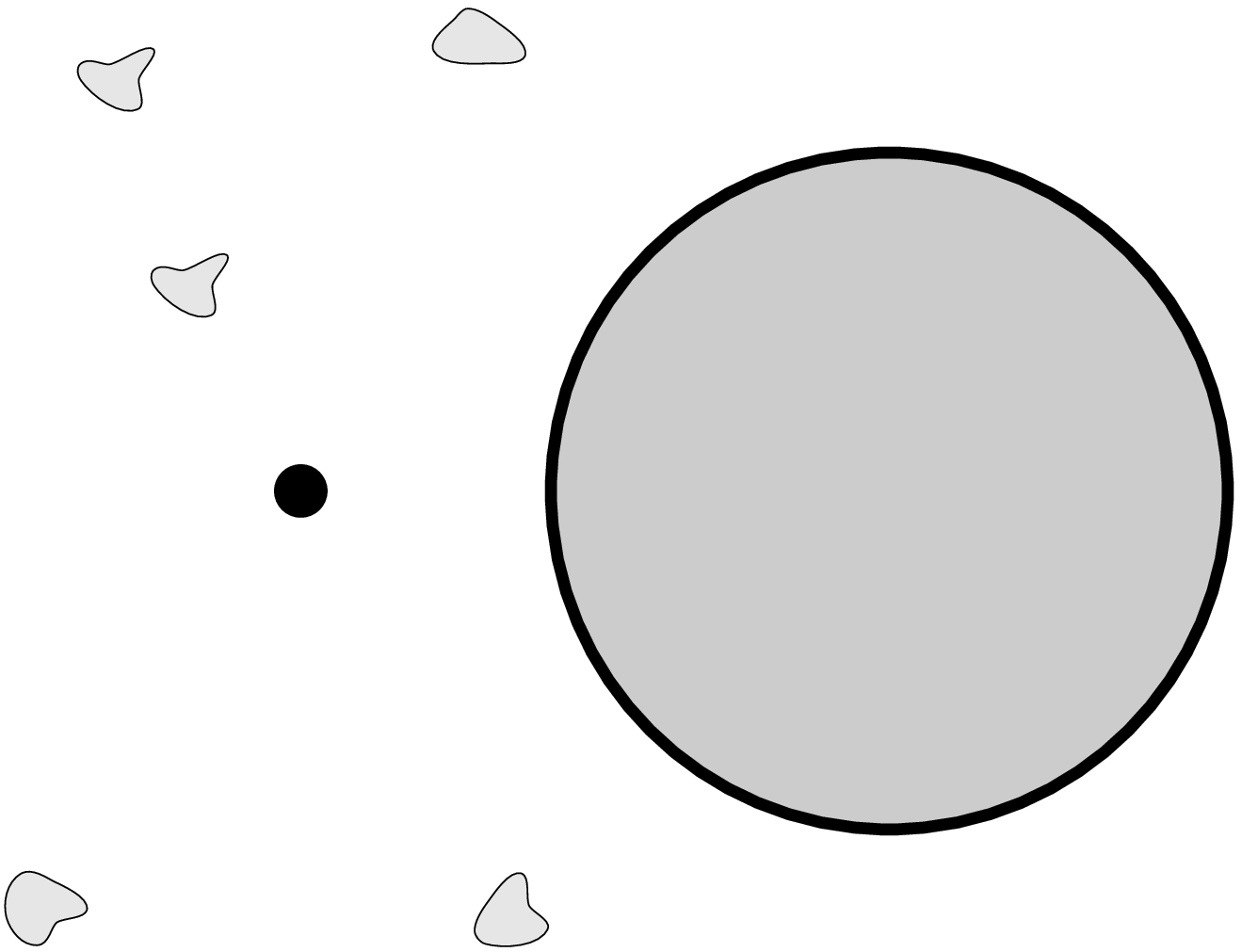}\hspace{0.5in}
\includegraphics[angle=-90,bb=-83 -27 295 263,width=2.4in]{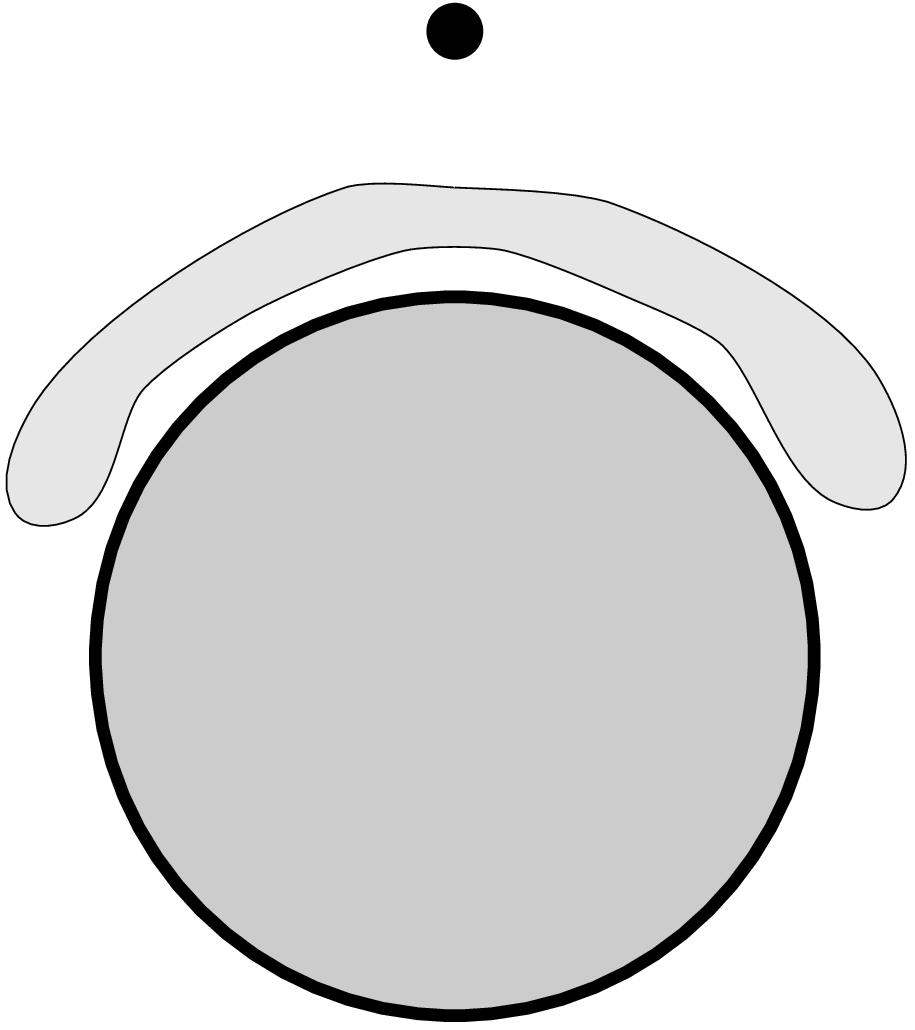}
\end{center}
\caption{In this figure we illustrate to possible distributions of the
helium-like ions consistent with the helium-like line fluxes we
measure.  In the first panel we illustrate the possibility that the
helium-like ions are in clumps distributed throughout the wind.  In
the second panel we illustrate the possibility that the helium-like
ions are near the surface of the companion star.}
\label{fig:clump_face}
\end{figure}

One possible explanation of our results is that the helium-like ions
exist primarily in dense clumps throughout the wind.  Dense clumps in the
wind of an HMXB have previously been invoked in the case of Vela~X-1
by \citet{sak99} in order to explain the strong fluorescence lines
observed from that system.  Another possible explanation is that
helium-like ions exist mainly near the surface of the star where the
density is higher, the ionization less, and the velocity lower than in
the outer parts of the wind \citep[see, e.g.,][]{lie01}.  Furthermore,
the part of a volume along the surface of the companion star with a
thickness significantly less than one stellar radius that is visible
during eclipse would subtend a solid angle significantly less than
$4\pi$ from the neutron star.  In Figure~\ref{fig:clump_face} we
illustrate both of these possibilities.
Definitive tests of these hypotheses would require calculations of the
line emission from detailed wind models and that is beyond the scope
of this work.  
However, we expect that our results, including the
remarkably small values of $\Omega\Delta{}v$ are consistent with
current expectations about the nature of winds in HMXBs.

\bibliographystyle{apj} 
\bibliography{hmxb}

\end{document}